\documentclass[usenatbib]{mn2e}

\usepackage{graphicx}
\usepackage{multirow}
\usepackage{natbib}
\def\citeN{\citet}
\def\cite{\citep}

\footnotesize
\newdimen\digitwidth    
\setbox0=\hbox{\rm0}
\digitwidth=\wd0
\catcode`!=\active
\def!{\kern\digitwidth}
\normalsize
\title[HTRU I: System configuration]{The High Time Resolution Universe Pulsar Survey I: System configuration and initial discoveries}
\author[M.~J.~Keith et al.]
{M.~J.~Keith$^{1}$\thanks{Email: mkeith@pulsarastronomy.net},
A.~Jameson$^{2}$,
W.~van Straten$^{2}$,
M.~Bailes$^{2}$,
S.~Johnston$^{1}$,
M.~Kramer$^{3,4}$, \newauthor
A.~Possenti$^5$,
S.~D.~Bates$^4$,
N.~D.~R.~Bhat$^{2}$,
M.~Burgay$^5$,
S.~Burke-Spolaor$^{2,1}$, \newauthor
N.~D'Amico$^{5}$,
L.~Levin$^{2,1}$,
P.~L.~McMahon${6,7}$,
S.~Milia$^{5,8}$,
B.~W.~Stappers$^4$
\\
$^1$ Australia Telescope National Facility, CSIRO, P.O. Box 76, Epping, NSW 1710, Australia\\
$^2$ Swinburne University of Technology, Centre for Astrophysics and Supercomputing Mail H39, PO Box 218, VIC 3122, Australia\\
$^3$ Max Planck Institut f\"ur Radioastronomie, Auf dem H\"ugel 69, 53121 Bonn, Germany\\
$^4$ University of Manchester, Jodrell Bank Centre for Astrophysics, Alan Turing Building, Manchester M13 9PL, UK\\
$^5$ INAF-Osservatorio Astronomico di Cagliari, localit\`a Poggio dei Pini, strada 54, I-09012 Capoterra, Italy\\
$^6$ Berkeley Wireless Research Center, University of California, Berkeley, CA 94704, USA\\
$^7$ Department of Electrical Engineering, University of Cape Town, Rondebosch, 7701, South Africa\\
$^8$ Dipartimento di Fisica, Universit\`a degli Studi di Cagliari, Cittadella Universitaria, 09042 Monserrato (CA), Italy \\
}
\date{}
\begin{document}

\maketitle
\newcommand{\setthebls}{
}

\setthebls

\begin{abstract} 
We have embarked on a survey for pulsars and fast transients using the 13-beam Multibeam receiver on the Parkes radio telescope.
Installation of a digital backend allows us to record 400~MHz of bandwidth for each beam, split into 1024 channels and sampled every 64~$\mu$s.
Limits of the receiver package restrict us to a 340~MHz observing band centred at 1352~MHz.
The factor of eight improvement in frequency resolution over previous multibeam surveys allows us to probe deeper into the Galactic plane for short duration signals such as the pulses from millisecond pulsars.
We plan to survey the entire southern sky in 42641 pointings, split into low, mid and high Galactic latitude regions, with integration times of 4200, 540 and 270~s respectively.
Simulations suggest that we will discover 400 pulsars, of which 75 will be millisecond pulsars.
With $\sim\!30\%$ of the mid-latitude survey complete, we have re-detected 223 previously known pulsars and discovered 27 pulsars, 5 of which are millisecond pulsars.
The newly discovered millisecond pulsars tend to have larger dispersion measures than those discovered in previous surveys, as expected from the improved time and frequency resolution of our instrument.

\end{abstract}


\section{Introduction}

Pulsars are fascinating celestial objects, with spin periods ranging from $\sim\!1.6$~ms to $\sim\!8.5$~s and magnetic fields that span more than five orders of magnitude.
Their study yields a better understanding of a variety of physics problems, from acceleration of particles in the ultra-strong magnetic field, to tests of gravity in the strong field regime.
Pulsars provide useful probes of their environments whether inside pulsar wind nebulae, the centres of
globular clusters or the Galactic Centre.
Discovery of these unique objects has continued unabated for more than 40
years, and the diversity in their properties means that, even though nearly 2000 pulsars,
including $\sim\!170$ millisecond pulsars (MSPs), are currently known, further discoveries are warranted.

In the past decade alone, surveys have uncovered a number of intriguing
objects. These include the pulsar with the fastest rotation period (in the
globular cluster Terzan 5; \citealp{hrs+06}), the double pulsar system \cite{bdp+03,lbk+04}
and its impressive tests of general relativity \cite{ksm+06}, the `missing link'
between the low-mass X-ray binaries and the MSPs \cite{asr+09}, rotating radio transients \cite{mll+06},
and young, energetic pulsars such as PSR~J1028--5819 \cite{kjk+08} with
their subsequent tie to gamma-ray pulsars and their emission properties
\cite{waa+10}. Furthermore, timing of MSPs
provides highly accurate parameters \cite{vbc+09} possibly leading
to the detection of gravitational waves in the near future \cite{hbb+09}.
Finally, the discovery of a burst of radio emission of unknown origin but
seemingly at cosmologically distances \cite{lbm+07} has motivated searches
for other such events.

Since the discovery of pulsars \cite{hbp+68}, searches have been split into targeted and untargeted surveys.
Targets of interest include, for example, globular clusters, supernova remnants, the Galactic Centre, and other regions known to be rich in pulsars.
Although targeted searches are highly efficient in their use of telescope time, it is impossible to predict the location of the majority of pulsars.
Therefore, untargeted searches over large areas of sky are the only way to significantly increase the number of known pulsars.

Since the early 1990s, large-area pulsar surveys with the Parkes telescope,
such as the low frequency Parkes Southern Pulsar Survey (PSPS; \citealp{mld+96}) and the higher frequency Parkes Multibeam Pulsar Survey (PMPS; \citealp{mlc+01}), Swinburne Intermediate Latitude Survey (SILS; \citealp{ebvb01}) and its extensions \cite{jbo+09} have led to the discovery of nearly 1000 pulsars.
The all-sky PSPS was highly
successful in finding MSPs \cite{lml+98}, but the low observing frequency limited the
survey volume, especially towards the Galactic plane.
The success of the later surveys was due to a combination of the higher
observing frequency, the low system temperature and the large field
of view provided by the 13-beam `Multibeam' receiver \cite{swb+96}.
These surveys had relatively coarse frequency resolution however, 
and this limited the searchable volume for short duration pulses from
MSPs and/or transient bursts.
For the PMPS this limit was due to the costs associated with 
replicating complex sets of analogue filters over 13 beams and the
processing power available in the late 1990s.

In the 14 years since the commissioning of the Multibeam receiver, there has been a great advance in the development of backend recording technology.
As pen chart recorders gave way to digital data analysis in the early 1970s,
so analogue signal processing equipment is now being replaced by digital 
systems that offer a more affordable way to obtain higher time and 
frequency resolution.
Increased frequency resolution allows for better correction of the dispersive effects of the interstellar medium, which, in combination with the higher time resolution, gives increased sensitivity to short duration pulses, such as those from MSPs.

Encouraged by these developments, we conceived the `High Time Resolution Universe' (HTRU) survey, motivated by two main drivers: to increase our understanding of the population of MSPs and to characterise the transient sky on timescales down to tens of microseconds.
These drivers led us to build a new digital backend system connected to the Parkes Multibeam receiver.
We use this new backend to give the HTRU survey four times the time resolution and eight times the frequency resolution of the PMPS.
This allows us to detect MSPs for which interstellar dispersion had hidden from previous surveys (Bates et al. in prep).
With the discovery of the radio magnetar PSR J1622--4950 \cite{hitrun_magnetar}, we have also shown that the transient nature of some exotic pulsars gives us good reason to re-survey areas of sky that are already well studied.

The HTRU survey intends to be an all sky survey for pulsars and short duration radio transients, with a strong focus on the lower Galactic latitudes, where we make most use of the higher frequency resolution for negation of interstellar dispersion.
A parallel effort is taking place at the Effelsberg radio telescope, covering the northern sky to a similar sensitivity, so as to provide a true all-sky survey at high time resolution.

In Section 2 of this paper we describe the survey strategy and the predicted results.
Section 3 outlines the observing setup in hardware and software.
Section 4 describes the analysis software that is being used to process the survey data.
In Section 5 we compute and measure the effective sensitivity limits of the survey.
Finally, in Section 6 we give the basic parameters of the first pulsars discovered in the HTRU survey.

\section{Simulation and survey strategy}
\begin{table}
\caption{\label{surv_params}
Survey parameters for the three parts of the HTRU survey.
}
\begin{center}
\begin{tabular}{l|lll}
Survey                      & High               & Mid                               & Low \\
\hline
\multirow{3}{*}{Region} & $\delta < +10^\circ$   & $-120^\circ < l$               & $-80^\circ < l$              \\
                        &                        & $l < 30^\circ$                 & $l < 30 ^\circ$              \\
                        &                        &$|b| < 15^\circ$                & $|b| < 3.5^\circ$            \\
$\tau_{\rm obs}$ (s)       & 270                    & 540                                   & 4300                                \\
$N_{\rm beams}$         & 443287                 & 95056                                 & 15990                               \\
$\tau_{\rm samp}$  ($\mu$s) & 64 &64&64\\
$B$ (MHz)              & 340 &340&340\\
$\Delta \nu_{\rm chan}$ (kHz) & 390.625& 390.625& 390.625 \\
$N_{\rm chans}$         & 870& 870& 870 \\
Data length (samples)   & $\sim\! 2^{22}$          & $\sim\! 2^{23}$                         & $\sim\! 2^{26}$                       \\
Data/beam (GB)          & 1.0                    & 2.0                                   & 16.0                                \\
Data/total (TB)         & 435                    & 190                                   & 250                                 \\
\end{tabular}
\end{center}
\end{table}

The HTRU survey of the southern sky is composed of three parts, outlined in Table \ref{surv_params}.
Other than the observation time and sky coverage, the observing parameters are identical for all three parts.
The {\bf low-latitude} component covers a thin strip of the Galactic plane, with Galactic latitude $|b| < 3.5^\circ$ and longitude $-80^\circ < l < 30^\circ$.
The 4200~s observation time is twice that of the PMPS, providing the most thorough survey of the inner Galactic plane to date.
The {\bf mid-latitude} component covers $|b| < 15^\circ$ and $-120^\circ < l < 30^\circ$.
Here the observation time is 540~s, allowing for a large area of sky to be covered quickly.
Finally, the {\bf high-latitude} component covers the entire sky south of declination $+10^\circ$ with 270~s integrations, excluding the region covered by the mid-latitude survey.
Broadly speaking, the aims of the low-latitude part is to discover faint pulsars deep in the Galactic plane, whereas the mid-latitude survey will find bright MSPs suitable for timing array projects.
Finally, the high-latitude survey will give us a snapshot of the transient sky at 64~$\mu$s resolution.

\begin{figure}
\begin{center}
\includegraphics[width=7cm]{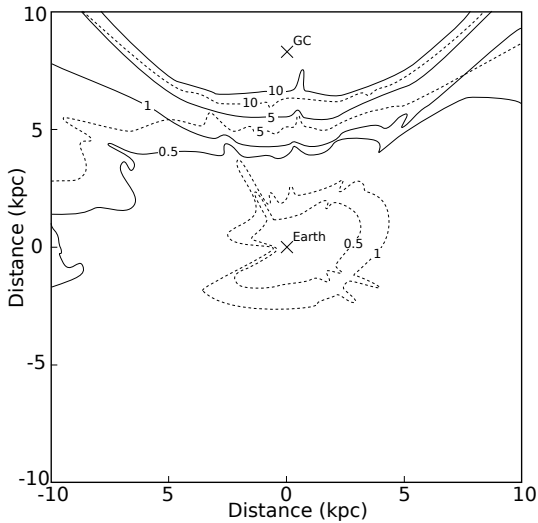}
\end{center}
\caption[]{
\label{galacticplot}
Contours of constant pulse broadening timescale for lines of sight at $b=0$ for a frequency of 1352~MHz.
The values for the PMPS are shown in dashed lines and for the HTRU survey in solid lines, with contours at 0.5, 1, 5 and 10~ms.
Scattering timescales and DMs were computed using the model of \citeN{cl02}.
The Earth and the approximate location of the Galactic Centre (GC) are shown with crosses.
}
\end{figure}

Detection of MSPs is limited by a combination of dispersion measure (DM) broadening in individual frequency channels, scattering in the interstellar medium and luminosity.
For sufficiently luminous MSPs, the PMPS was predominantly limited by dispersion broadening in the 3~MHz channels, rather than by scattering.
The increased frequency resolution of the HTRU survey means that the broadening is now dominated by scattering in essentially all directions.
Figure \ref{galacticplot} shows contours of constant pulse broadening for the PMPS and HTRU surveys, taking into account both DM and scattering.
Even in the worst case, towards the Galactic Centre, the limiting distance of the HTRU survey is greater than twice that of the PMPS, and significantly more in all other directions.
This implies that the mid-latitude survey, with an integration time only one quarter that of the PMPS, should still discover pulsars that lie in previously surveyed regions.
The deeper low-latitude survey will be able to penetrate further into the Galactic plane for both MSPs and longer period pulsars.

In order to determine the discovery potential of the HTRU survey, we ran simulations of the pulsar population based on the model presented in \citeN{lfl+06} using the {\sc psrpop}\footnote{http://psrpop.sourceforge.net} software.

For the normal pulsar population, the input parameters to the model were as follows:
\begin{itemize}
\item A log-normal spin period distribution, with mean $10^{2.71}$~ms and sigma $10^{0.34}$~ms.
\item A simple power-law for the luminosity distribution, with index $-0.59$ and low luminosity cut off at 0.1~mJy~kpc$^2$.
\item An exponential function for the height above the Galactic plane, with scale 0.33~kpc.
\item The radial distribution of \citeN{yk04}.
\item A total of 28000 pulsars, such that the simulated detection of the PMPS matched closely to the true discovery rate.
\end{itemize}

The number of known Galactic MSPs is small and many previous surveys have been restricted by interstellar dispersion, making simulation of the Galactic population of MSPs more uncertain and indeed this population was not considered by \citeN{lfl+06}.
To produce simulations consistent with observed detection rates, we made the following modifications to the model:
\begin{itemize}
\item The period distribution was based on that of the known Galactic plane MSPs.
\item A larger scale height of 0.5~kpc was used, to better match the known MSP population \cite{cc97}.
\end{itemize}

Along with the three parts of the HTRU survey, we also simulated the PSPS, PMPS and SILS surveys.
The results are summarised in Table \ref{simulations}.
In total, the HTRU surveys should discover nearly 400 pulsars, of which 75 will be MSPs, which would represent a doubling of the known population of MSPs.
Final results from the survey will also allow us to obtain a better description of the radial and velocity distributions of MSPs.

\begin{table}
\caption{\label{simulations}
Simulated results for the mean number of total pulsar detections and new discoveries for three previous surveys and the three components of the HTRU survey.
The numbers are split the simulation of `normal' pulsars and `MSPs' as described in the text.
The numbers in parentheses are the actual numbers of pulsars found in each of the complete surveys, where
we defined a pulsar as an MSP if it had a period of less than 30~ms and a period derivative less than $10^{-17}$.
}
\begin{center}
\begin{tabular}{l|llll}
Survey & \multicolumn{2}{c}{Normal} & \multicolumn{2}{c}{MSPs} \\
       & \multicolumn{1}{c}{Total}  & \multicolumn{1}{c}{New}   & \multicolumn{1}{c}{Total}  & \multicolumn{1}{c}{New}    \\
\hline
PSPS   & 281 (279)  & --  (89) & 20 (19) & -- (17) \\
PMPS   & 1008 (1047)& 876 (721) & 19 (20) & 18 (21) \\
SILS   & 137 (158)  & 57  (61)  & 10 (11) & 6 (8)   \\
HTRU & & & &\\
~~Low-lat& 957        & 260       & 51      & 33      \\
~~Med-lat& 831        & 52        & 48      & 28      \\
~~High-lat& 783        & 11        & 65      & 13      \\
\end{tabular}
\end{center}
\end{table}

\section{Observing system}
Observational data are acquired using the Parkes 21-cm Multibeam receiver \cite{swb+96} and a new digital backend system, the details of which are described in the following sections.

\subsection{Analogue systems}
\begin{figure}
\begin{center}
\includegraphics[width=7cm]{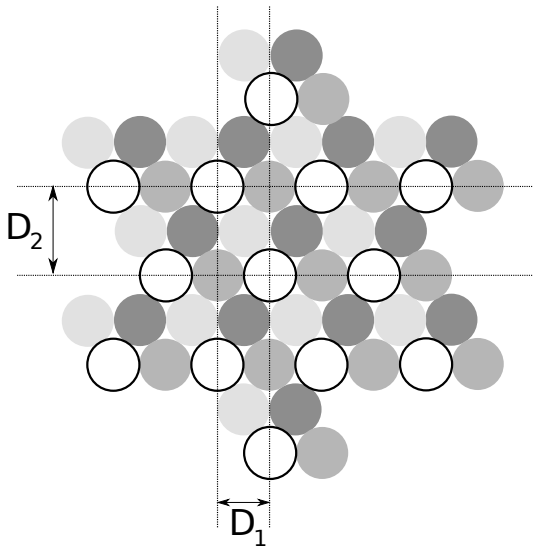}
\end{center}
\caption[]{
\label{multibeam}
The beam pattern of the Multibeam receiver.
The black circles indicate the 14\arcmin~FWHM of each of the 13 beams of the receiver.
The separation D$_1$ is also equal to this 14\arcmin~FWHM, and ${\rm D}_2=\sqrt{3}{\rm D}_1$
The filled circles show three further pointings interleaved to make up the the shape from which the entire survey is tessellated.
}
\end{figure}

The Parkes Multibeam receiver consists of 13 feeds at the prime focus of the Parkes 64~m antenna, organised as a central feed surrounded by two hexagonal rings, with beam pattern on the sky as shown in Figure \ref{multibeam}.
Our measurement of the system temperature of the central beam -- 23~K at 1400~MHz -- is slightly higher than the value quoted in \citeN{mlc+01}, which may be due to the refurbishment of the analogue electronics in 2006 (J.~Reynolds 2010, private communication).
The central feed has a symmetric beam pattern, and the feeds further from the prime focus have a slight ellipticity and gain degradation.
The details of the receiver and telescope setup are listed in Table \ref{telrx}.
For each beam of the receiver, both polarisations are down-converted from 1182-1582~MHz to 0-400~MHz.

Unfortunately, the high-frequency part of this band is badly affected by interference from the Thuraya 3 geostationary communications satellite, 
which transmits a space-to-earth signal in the 1525-1559~MHz band.
We have installed low-pass hardware filters after the receiver that restrict the upper end of our observing band to 1522~MHz;
therefore, the effective observing band is 340~MHz wide, centred on 1352~MHz

\begin{table}
\caption{\label{telrx}
Details of the Multibeam receiver for the central feed, the inner ring, and outer ring of feeds,
taken from \citeN{mlc+01}.
}
\begin{center}
\begin{tabular}{llll}
Beam & Centre & Inner & Outer\\
\hline
Telescope Gain (K Jy$^{-1}$) & 0.735 & 0.690 & 0.581 \\
Half-power beam width (\arcmin) & 14.0 & 14.1 & 14.5 \\
Beam ellipticity & 0.0 & 0.03 & 0.06 \\
Coma lobe (dB) & none & 17 & 14 \\
\end{tabular}
\end{center}
\end{table}

\subsection{The BPSR backend System}
The Berkeley-Parkes-Swinburne Recorder (BPSR) system consists of 26 digital spectrometers connected to 13 server-class workstations that format the data and write it to disk.
The digital spectrometers are based on the IBOB\footnote{Internet Break-Out Board} platform developed by the CASPER\footnote{Center for Astronomy Signal Processing and Electronics Research} group at the University of California, Berkeley, described in detail in \citeN{mcm08}.

Each IBOB consists of two analogue-to-digital converters and two polyphase filters implemented on an FPGA\footnote{Field Programmable Gate Array}.
Each pair samples orthogonal polarisations at 800~MHz using 8 bits per sample, then divides each signal into 1024 spectral channels of width 390~kHz.
Each spectral channel is detected and integrated over 25 samples to yield the output time resolution of 64~$\mu$s.
The 32-bit output of the spectrometer is linearly scaled before decimation to 8 bits; the 8-bit data are written to a single UDP\footnote{User Datagram Protocol} packet with a 2048~byte payload containing 1024 channels times 2 polarisations.  Each packet is sent to a 10~Gb/s Ethernet (10GbE) interface connected directly to a Dell PowerEdge 1950 server via a CX4 cable.

On the server, data are collected into 10~s blocks and further processed.  Both polarisation streams of each channel are summed and
then normalised by subtracting the mean and dividing by the standard deviation (as measured in the first 10~s block of each observation).
The resultant data are then decimated to 2 bits per sample and written to disk at a rate of 32~Mb/s.

\subsection{Data collection}
Each server in the BPSR cluster has sufficient disk space to buffer $\sim\!3$~days of observations.
For the HTRU survey, we have chosen to archive the data to magnetic tapes before removing the data from the buffer, allowing for continuous observing.
One copy is written to tape at Parkes and another is streamed via a dedicated 1~Gb/s fibre link to the supercomputer located at Swinburne University.
Here the data are also written to magnetic tape and, if there is sufficient free capacity on the supercomputer, run through the survey data processing pipeline in close to real-time.

In addition to the astronomical signal, the BPSR backend also records a header file and a collection of auxiliary data files for each beam.
The header file contains information about the observation in an XML\footnote{eXtensable Markup Language, http://w3.org/TR/REC-xml/} format as described in Section \ref{psrxml}.
The auxiliary files contain bandpass and low resolution time series snapshots, as described in Section \ref{aux_files}.
The raw time series is stored in the {\sc sigproc}\footnote{http://sigproc.sourceforge.net} `filterbank' data format, which consists of a binary header followed by continuous raw data.

For the HTRU survey, all data are recorded using two bits of precision and with four samples packed per byte, with the first sample in the lowest significant bits of the byte.
The data are ordered first by frequency then by time, each time sample being stored as a consecutive 256 byte word.
The raw data are preceded by a 345 byte header, used for backwards compatibility with the {\sc sigproc} software, however more information is available in the XML header file.
These data are then packed with the header and auxiliary files into a single GNU tarball per beam.

\subsection{Psrxml header format}
\label{psrxml}
The psrxml header is a format-independent XML header file that describes both the observational properties of the recorded data and the format of the actual data itself.
The psrxml header file is stored separately to the raw data, allowing trivial indexing of header data for very large data files, such as those produced by the BPSR backend.

The psrxml data format is intended to be suitable for long term archival of pulsar data for future uses.
Features of note are:
\begin{itemize}
\item Checksumming of data blocks: As the data are written to disk, each block of data (usually a few hundred megabytes) is passed through the SHA-1\footnote{Secure Hash Standard, Federal Information Processing Standards Publications, 180-1, 1995} hash algorithm and the hash value stored in the header file. This allows for the data to be checked for errors and consistency at a later time. Archival data are valuable: This check-summing ensures that data extracted from an archive are identical with the data recorded at the telescope.
\item Format and machine independence. The psrxml header file is written in a well-defined format, typically encoded in standard 7-bit ASCII\footnote{American Standard Code for Information Interchange, American National Standards Institute, X3.4-1986, 1986}. This means that the header will be readable on almost any computer that exists or is likely to exist. Since the header completely describes the data format, in terms of bit and byte ordering, and which bytes to read, one can therefore read the raw data on any computer with the appropriate software.
\item Human readable. XML documents can be easily read using a standard text editor. With the use of XSL\footnote{eXtensable Stylesheet Language}, viewing a psrxml header file in a web browser displays the data in a pleasant format without the use of any specialist software. This does not change the file in any way, so the file is readable in ideal formats by both machine and human.
\item Extendable. Since the header is written in XML, it is trivial to add new XML tags to the file in a safe manner, storing either fields specific to a certain project, or extending the data format to new generic uses.
\end{itemize}

\subsection{Auxiliary files}
\label{aux_files}
In addition to the header and data files, data generated for on-line monitoring of the system are stored for future use.
This consists of the passband and time series measured in the pre-normalised, dual-polarisation, 8-bit data samples, completely averaged in frequency.

\subsection{Data tracking}
At the completion of every observation, a copy of the psrxml header file is kept on disk and indexed in a book keeping database.
The database links the observation against the planned observations and marks that observation as completed, removing that pointing from the pool of pointings from which observing schedules are selected.

\section{Processing}
Processing of the survey is currently being carried out at the Jodrell Bank Centre for Astrophysics and the Swinburne Centre for Astrophysics and Supercomputing.

The analysis of our data is limited by the available computational power, therefore re-processing of the data with future computer hardware will likely yield additional results.
We follow a typical pulsar search workflow as described, for example, in \citeN{lk05}.

\subsection{Software Pipeline}
For our survey processing we have developed the {\sc hitrun} processing pipeline.
{\sc hitrun} is designed to be a flexible system that can process both BPSR data and legacy 1-bit data formats without modifying the software.
This flexibility, as well as being a useful tool, allows us to easily test the software on legacy data and against legacy codes.

The {\sc hitrun} pipeline is composed of several stages: Radio frequency interference (RFI) removal, dedispersion, time series analysis, candidate sorting, candidate folding and optimisation.
Here, we outline the algorithms used for each step of the process.

\subsubsection{RFI removal}
\label{rfizap}
RFI removal consists of two tasks, removal of `bad' spectral channels, and removal of `bad' time samples.

First, we remove spectral channels that are affected by periodic RFI.
Initially, we have a list of channels that are known to have poor data quality, and these are always removed.
Then, we Fourier analyse each of the channels individually, using the Fourier time series analysis described below.
If any signals are detected with a signal-to-noise ratio (SNR) of greater than 15, we reject that channel.

\begin{figure}
\begin{center}
\includegraphics[width=8.5cm]{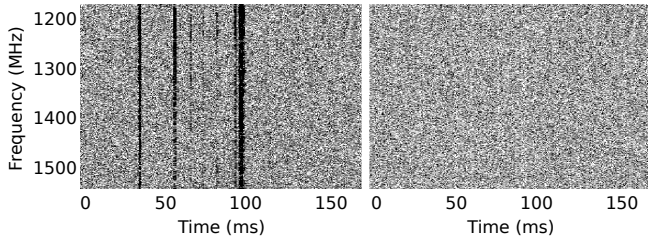}
\end{center}
\caption[]{
\label{zap}
The intensity of recorded signals over a short time as a function of frequency, before (left) and after (right) RFI removal.
The RFI can be seen as a series of vertical stripes, which are replaced with random noise by the RFI excision routine.
}
\end{figure}

Next, we sum the data in frequency, producing a single time series at DM $=0$.
This time series is then searched for bursts of emission, and if found, those samples are removed from the time series.
To remove longer duration RFI bursts, we sum together 2, 4, 8 16 and 32 adjacent samples and repeat the search.
Any time samples that are removed in this process are then removed from the original data file, replaced by samples that are drawn from a random population statistically indistinguishable from the data.
An example of this RFI removal is shown in Figure \ref{zap}, where a series of short-duration wide-band signals have been removed from the data.
It should be noted that this process will remove low-DM astrophysical signals of sufficient intensity and width.
With our \mbox{5-$\sigma$} threshold for removal, astrophysical signals of maximum recordable intensity will be removed if their DM is less than $0.12(W/64\mu s)$ cm$^{-3}$pc, where $W$ is the approximate width of the pulse.

\subsubsection{Dedispersion}
Since the DM of any signal is not known a priori, we have to search many trial DM values.
We must therefore transform the single time series with many frequency channels into many frequency-summed time series with different dispersion delays.
For our survey, the processing time is largely dominated by the dedispersion time, making it important to choose the most efficient algorithm possible.

The most efficient dedispersion algorithm known is the `tree' dedispersion approximation of \citeN{tay74}.
This, however, assumes a linear dependence of delay on frequency, which is a good approximation for small fractional bandwidths.
For many recent surveys with large fractional bandwidths this approximation is inappropriate, and work-arounds have been used.
In the PSPS, a two-step approach was used where smaller sub-bands were dedispersed using the tree algorithm, before a final dedispersion step to combine the sub-bands \cite{mld+96}.
In addition to this, the PMPS used a linearisation approach where the data were padded with empty channels in a manner such that the dispersion sweep could be considered linear \citeN{mlc+01}.
For the HTRU survey we have decided not to use the tree algorithm as the cost of linearising the data outweighed the gain from the more efficient algorithm.
Instead, we have chosen to implement an efficient $\nu^{-2}$ dedispersion algorithm.

Much of the cost of dedispersion is due to the large volume of data that must pass through the system, therefore we compact the data as much as possible.
Since we sum 1024 2-bit values per output sample, there are 4096 possible output levels.
Our processing systems perform 64-bit integer arithmetic, so we can pack 4 samples per 64-bit word, giving a factor of 4 processing throughput increase.
By dividing the output by 16, the number of levels is reduced to 256, with little sensitivity loss.
We therefore only need to write out 8-bits per sample, keeping the input/output costs to a minimum.
Additionally, we make use of modern multi-core processors with multi-threaded code performing 4 dedispersion threads simultaneously.
In future, it will likely be possible to gain additional performance benefits with technologies such as graphics processing unit co-processing.

\subsubsection{Time series analysis}
By this stage of the processing we have a number of dedispersed time series on disk, which have to be searched for pulsar signals.
{\sc hitrun} provides both time-domain and frequency-domain methods.

In the frequency domain we take the power spectrum of the data using a fast Fourier transform.
We then remove the strong red-noise component of the spectrum at frequencies below about 10~Hz by subtraction of a running mean and division by a running variance.
This also has the effect of normalising the spectrum.
The resulting power spectrum is then searched for signals with a signal-to-noise ratio of greater than 6.
To reclaim power in the harmonics of narrow pulses, we also search spectra which have had 2, 4, 8 and 16 harmonics summed.
Potential signals from each DM trial are written out for later sorting, as described below.

In the time-domain we conduct a search with similar methodology to that reported in \citeN{bb10}. In brief, we search for bright, single pulses by collating all points above 6-$\sigma$ from the mean, after which the detections from a single pointing are compared across all the beams to identify sky-localised events. Details of the single-pulse-specific analysis and the results of the time-domain search will be reported in a future paper.

\subsubsection{Candidate sorting}
Candidate sorting is the process of collecting the Fourier detections from each DM step and combining those into a manageable number of candidate pulsars for further analysis.
This is done by grouping detections with the same frequency into a single candidate, and by combining candidates that are harmonically related.
In {\sc hitrun}, this is done using software from {\sc pulsarhunter}\footnote{http://www.pulsarastronomy.net/wiki/Software/PulsarHunter}.

\subsubsection{Candidate folding and optimisation}
We then fold the data, producing an averaged pulse profile for each candidate.
Folding is a coherent technique unlike the incoherent harmonic summing in the Fourier domain.
{\sc hitrun} utilises the {\sc dspsr}\footnote{http://dspsr.sourceforge.net} software to fold each candidate, producing profiles for each of 16 frequency bands and 32 sub-integrations.

Then the {\sc psrchive} package \cite{hvm04} is used to sum the folded sub-integrations and periods with a series of trial periods and DMs around the values returned from the Fourier domain search.
The optimised period and DM, i.e. that which give highest folded SNR, are then used to make the final diagnostic plots for the candidate.

\subsection{Selection of candidates}
The goal of the data processing is to select a number of likely pulsar signals for re-observation.
To achieve this, we take the optimised candidates and the associated diagnostic plots and view them with graphical candidate selection tools, such as {\sc JReaper} \cite{kel+09}.
These tools reduce the large number of candidates to a series of points on a scatter chart, usually with period and SNR on the axes.
Selecting these points presents the user with the full candidate details and diagnostic plots, at which point the user must exercise judgement on whether the candidate should be flagged for further follow up.

Since manual candidate selection is a time consuming and potentially error prone process, we are investigating the potential for using artificial neural network (ANN) algorithms to pre-select promising candidates for viewing \cite{emk+10}.
An ANN is a mathematical construct, consisting of a set of nodes that are connected to form a decision tree that, based on a given input vector, derives a conclusion about the validity of a hypothesis.
In our case, we provide the ANN with a vector of `scores', formed from analysis of the diagnostic plots, e.g. the agreement of the observed SNR-DM curve of a candidate with theoretical predictions as measured by a $\chi^2$ estimator.
The more types of scores that are used, the larger (usually) the chance to differentiate between pulsar and non-pulsar signals, however the more known pulsar signals are required to `train' the ANN \cite{emk+10}.
Indeed, our ANN is poor at selecting binary or millisecond pulsars, because of the limited size of training set available for these classes of pulsar.
The effectiveness of this technique can be seen in the discovery of PSR J1701--44, which was identified as a candidate by the ANN.
The SNR of the candidate was rather low, at only 7, and would quite probably have been missed by a graphical interface reliant upon a human to select the candidate manually.

Although the ANN is limited in the types of pulsars it will select, it does provide a thorough analysis of all the candidates free from human error.
We find that adding the ANN to our existing candidate selection effort gives a beneficial additional way to select candidates, with little cost in terms of required computing power.
Direct inspection of candidates is vital, however, and remains the primary way that pulsars have been discovered in the HTRU survey.

\section{Survey Sensitivity}
Knowledge of the minimum detectable flux density of the survey is vital for determining the survey effectiveness.
Determining this limiting flux density is not trivial, since the survey is not uniformly sensitive, even over a small region of sky, nor to all pulsars.
One option is to compute a theoretical mean sensitivity to some standard `typical' pulsar, as done by other large pulsar surveys.
This gives a value that can be easily compared to other similar surveys.

To measure the actual sensitivity of the survey, we can look at re-detections of known pulsars.
Since published values exist for the position and flux density of thousands of pulsars, we can cross match those positions with our survey pointings.
We then compute the theoretical SNR value for that pulsar given the offset in the beam position and compare that to the output of the search pipeline.
We consider each of these in in turn below.

\subsection{Theoretical sensitivity}
\label{theoretical_sens}

The radiometer equation can be used to compute the fundamental limiting flux density for the centre of the central beam of a given survey observation.
\begin{equation}
S_{\rm min} = \frac{\sigma (T_{\rm sys} + T_{\rm sky})}{G \sqrt{2 B \tau_{\rm obs}}},
\label{smin}
\end{equation}
where $\sigma$ is the minimum acceptable SNR, $T_{\rm sys}$ is the system noise temperature, $T_{\rm sky}$ the sky noise temperature, $G$ is the system gain, $B$ is the observing bandwidth and $\tau_{\rm obs}$ is the integration time.
$T_{\rm sky}$ is only a significant contribution to the system noise for low Galactic latitudes and has a strong dependence on the area of sky that is observed.
The value of $T_{\rm sky}$ at 1352~MHz for the survey region is extrapolated from the \citeN{hks+81} sky map at 408~MHz.
For Equation~\ref{smin} we use the mean values listed in Table~\ref{tsky}, however it should be noted that $T_{\rm sky}$, and therefore $S_{\rm min}$, varies considerably across the sky.
Here we use $\sigma=8$, $T_{\rm sys}=23$~K and $G=0.735$ to compute a $S_{\rm min}$ of 0.20, 0.47 and 0.61~mJy for the low, mid and high latitude portions of the survey respectively.
However as described below, the practical limiting flux density is lower due to the narrow pulse width of most pulsars.

Our Fourier based detection algorithm is not uniformly sensitive to all DMs, pulse periods and pulse shapes.
This is somewhat difficult to quantify, however we follow the approach of \citeN{mlc+01} by modelling the frequency response of the pulsar as a uniform train of impulses separated by frequency $f_p=1/P$.
This train of pulses has amplitude $1/S_{\rm min}$, to which we multiply a series of functions representing the various transformations that the data undertake.

We can model each pulse as a Gaussian with width $W_{50} = 0.05P$, so that the frequency response is multiplied by the Gaussian
\begin{equation}
F(f) = \exp{\left(-\frac{(\pi f W_{50})^2}{4 \ln(2)}\right)}.
\end{equation}

The pulses are also convolved with the effect of interstellar dispersion, which can be modelled as another Gaussian with width $\tau_{\rm DM}$, the dispersion delay.
Finally we must account for the finite sampling interval, $\tau_s$, which multiplies the amplitudes by
\begin{equation}
F(f) = \left|\frac{\sin{(\pi f \tau_s)}}{\pi f \tau_s}\right|.
\end{equation}

\begin{table}
\caption{
\label{tsky}
The maximum, minimum and mean sky noise temperature over the three survey regions.
Note that the values for the mid-latitude survey exclude the low-latitude survey region.
}
\begin{center}
\begin{tabular}{llll}

Survey & Mean (K) & Max (K) & Min (K) \\
\hline
Low & 7.6 & 36.0 & 1.6  \\
Mid & 2.5 & 9.1& 0.6 \\
High & 1.0 & 2.4 & 0.6\\
\end{tabular}
\end{center}
\end{table}

\begin{figure}
\includegraphics[width=8cm]{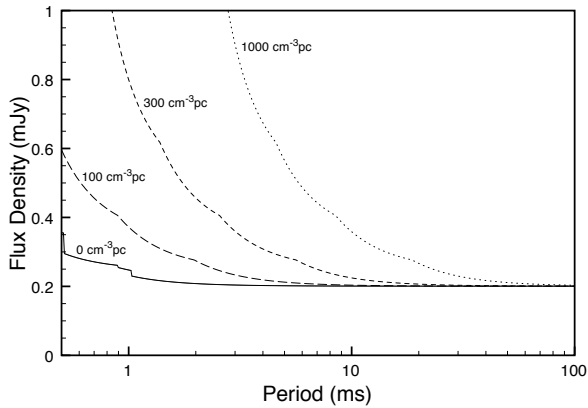}
\caption[]{
\label{med_sensplot}
The mean limiting flux density for the mid-latitude survey shown as a function of pulsar period for various values of DM.
Values shown here assume a pulse duty cycle of $5\%$.
}
\end{figure}

\begin{figure}
\includegraphics[width=8cm]{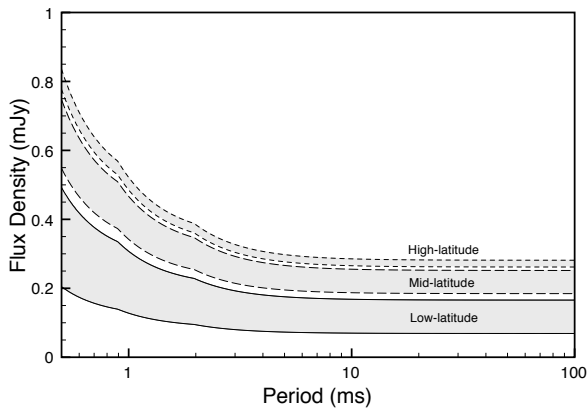}
\caption[]{
\label{sens_compare}
Comparison of the sensitivity of the three survey regions for a pulsar with a DM of 100~cm$^{-3}$pc.
Each region has two curves, showing the values for the minimum and maximum value of $T_{\rm sky}$ in the survey region.
}
\end{figure}

The final SNR is computed after summing $N$ harmonics, where $N \in \{1,2,4,8,16\}$.
Figure \ref{med_sensplot} shows the minimum detectable flux density as a function of pulsar period and DM for the mid-latitude survey.
It can be seen that the HTRU survey is sensitive to MSPs out to DMs of a few hundred cm$^{-3}$pc.
The effects of $T_{\rm sky}$ and the differing observing time for the three surveys can be seen in Figure \ref{sens_compare}, which shows the sensitivity curves for the three surveys at the minimum and maximum $T_{\rm sky}$ in that survey region.

The mean sensitivity of the survey is reduced by a number of factors.
Firstly, the outer beams of the receiver are less sensitive than the centre beam, and the sensitivity of each beam reduces towards the edges.
This means that the sky coverage is not uniform, with the mean sensitivity over the mid-latitude survey of about 0.25~mJy.

\subsection{Re-detection of known pulsars}
A total of 290 previously known pulsars lie within one beamwidth of one of the HTRU survey pointings collected to date.
For each pulsar we obtained the flux density and its error from the pulsar catalogue\footnote{http://www.atnf.csiro.au/research/pulsar/psrcat/}, a value of $T_{\rm sky}$ at the pulsar position, and the offset between its position and the centre of the relevant beam.
We then computed the expected SNR for each pulsar following the process outlined in Section \ref{theoretical_sens}.

In all, we detected 223 of the 290 pulsars.
For the 67 pulsars we failed to detect, 5 observations were ruined by broad-band RFI, 54 pulsars were not detected above a SNR of 10 even after we folded observations at the known pulsar ephemeris.
This leaves 8 pulsars that are seen in folded observations with a SNR of greater than 10 but were not detected in our search procedure.
Each of the the 8 pulsars had a spin period above 500~ms, and so the Fourier SNR may well have been reduced by a larger red-noise component in the data, or the pulsar period falling between Fourier bins.
The loss of pulsar detections due to RFI is an uncontrollable aspect of search observations.
Although we have applied the techniques described in Section \ref{rfizap}, they are unable to remove this type of broad-band RFI.
Future re-processing using the technique of \citeN{ekl09} may be possible, however this is currently unfeasible as it would increase the time for processing the survey data by more than a factor of ten.

\begin{figure}
\includegraphics[width=8cm]{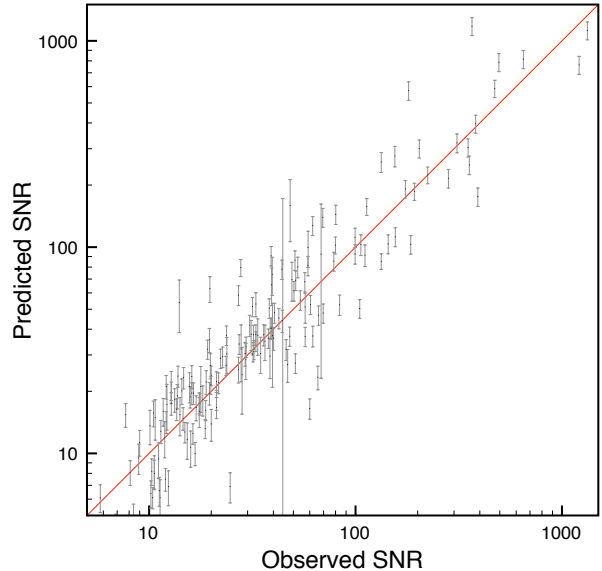}
\caption[]{
The predicted SNR for 223 known pulsars re-detected in the HTRU survey, shown as a function of the observed SNR.
Errors shown are from the published error in the pulsar flux density.
The line shows the expected 1:1 correlation.
\label{snrratios}
}
\end{figure}

Figure \ref{snrratios} shows the expected versus actual SNR for the 223 detected pulsars.
Although the correlation between the predicted and observed values is good, the predicted values are, on average, somewhat higher than the observed values.
This can be caused by either sensitivity loss in our observing system, or error in the catalogued flux densities.
First, we checked that the backend was not introducing any loss by collecting data simultaneously with BPSR and the Parkes Digital Filterbank 3 (PDFB3).
These observations show that the BPSR and PDFB3 performance are equivalent and we therefore consider it unlikely that the backend is introducing any losses.
Hobbs (private communication), has shown that the catalogue flux densities appear to be systematically higher than those obtained in interferometric observations.
This bias may be due to the tendency to publish only the data with the highest SNR, for example when scintillation boosts the apparent flux density.
Given our ability to detect the majority of the pulsars that are in our search area, we are relatively confident that the system is performing to specification.

\section{New pulsar discoveries}
\label{results}
\begin{table}
\caption{
Basic parameters for the 27 pulsars discovered in the HTRU survey to date.
Pulsars for which no full timing solution is published have been assigned a
temporary name containing only two digits of declination.
\label{psr_tab}
}
\begin{center}
{\footnotesize \tt 
\begin{tabular}{lrrlll}
\hline
PSR J & \multicolumn{1}{c}{$l$} & \multicolumn{1}{c}{$b$} & Period & DM  \\
 & \multicolumn{1}{c}{(\degr)} &\multicolumn{1}{c}{(\degr)} & (s) & (cm$^{-3}$pc) \\
\hline
1125$-$58$^1$ & 291.8 & 2.6 & 0.003102 & 125 \\
1330$-$52 & 308.8 & 9.6 & 0.648104 & 149 \\
1433$-$50 & 318.9 & 9.1 & 1.017495 & 98 \\
1442$-$51 & 320.1 & 7.7 & 0.732061 & 97 \\
1517$-$46 & 327.38 & 9.22 & 0.88661 & 139 \\
\\
1612$-$58 & 327.0 & $-$5.0 & 0.615520 & 172 \\
1623$-$4950$^2$ & 333.9 & $-$0.1 & 4.325800 & 881 \\
1625$-$49 & 334.6 & 0.0 & 0.355856 & 717 \\
1627$-$59 & 327.3 & $-$7.3 & 0.354239 & 92 \\
1648$-$36 & 347.1 & 5.8 & 0.212316 & 222 \\
\\
1650$-$60 & 328.2 & $-$10.2 & 0.583771 & 106 \\
1701$-$44 & 342.3 & $-$1.3 & 0.755535 & 413 \\
1705$-$43 & 343.3 & $-$1.5 & 0.222561 & 184 \\
1705$-$61 & 328.7 & $-$12.2 & 0.808546 & 87 \\
1708$-$35$^1$ & 350.4 & 3.2 & 0.004505 & 146 \\
\\
1731$-$18$^1$ & 6.9 & 8.2 & 0.002345 & 106 \\
1753$-$38 & 352.3 & $-$6.4 & 0.666804 & 168 \\
1754$-$24 & 4.9 & 0.8 & 2.090263 & 789 \\
1801$-$32$^1$ & 358.9 & $-$4.5 & 0.007454 & 177 \\
1802$-$33 & 357.6 & $-$5.5 & 2.461052 & 254 \\
\\
1803$-$33 & 358.0 & $-$5.6 & 0.633412 & 171 \\
1810$-$01 & 27.0 & 8.6 & 0.744976 & 135 \\
1811$-$24$^1$ & 7.1 & $-$2.5 & 0.002661 & 60 \\
1811$-$49 & 344.2 & $-$14.2 & 1.432704 & 49 \\
1812$-$27 & 4.0 & $-$4.6 & 0.236983 & 105 \\
\\
1814$-$05 & 23.8 & 5.8 & 1.014405 & 119 \\
1854$-$15 & 19.0 & $-$8.0 & 3.453121 & 157 \\

\hline
\multicolumn{5}{l}{
\rm $^1$ Timing parameters in Bates et al. (in prep).
}\\
\multicolumn{5}{l}{
\rm $^2$ Timing parameters in \citeN{hitrun_magnetar}.
}
\end{tabular}
}
\end{center}
\end{table}

In this paper we present the results of the analysis of the first $\sim\!30 \%$ of the mid-latitude part of the survey.
Results so far include:
\begin{itemize}
\item Discovery of 5 MSPs, all of which are in binary systems. Full details of these pulsars will be reported in Bates et al. (in prep).
\item Discovery of the radio pulsar with the highest known magnetic field which appears to be a radio-loud magnetar, as reported in \citeN{hitrun_magnetar}.
\item Discovery of 21 pulsars with periods between 0.2 and 3.5~s and
moderate to high DMs. These are likely `standard' solitary radio pulsars
but full timing solutions still need to be obtained.
\end{itemize}
The Galactic coordinates, spin period and DM for each of these pulsars is given in Table \ref{psr_tab}.

\subsection{Comparison to PMPS}
\begin{figure}
\includegraphics[width=8cm]{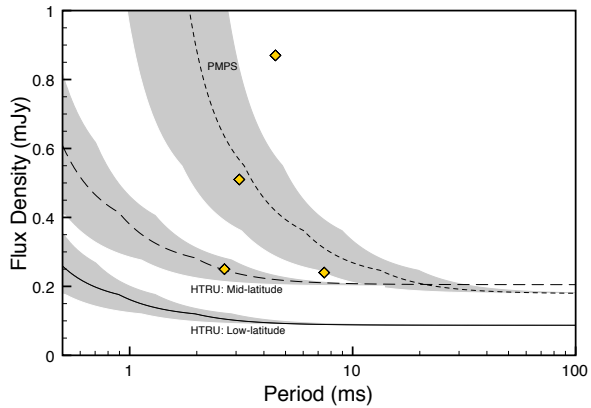}
\caption[]{
\label{mspsensplot}
Comparison of the Mid and Low latitude parts of the HTRU survey and the PMPS.
The lines give the sensitivity limit for a DM of 100~cm$^{-3}$pc.
The shaded areas show the range of sensitivities for DMs between 50 and 150~cm$^{-3}$pc.
Note that the limiting flux density for the Mid-latitude survey used for this curve is computed using the mean $T_{\rm sky}$ for the PMPS survey
area.
Also shown as diamonds, are the periods and estimated flux densities for the four MSPs discovered in HTRU survey that lie in the PMPS region.
}
\end{figure}

Figure~\ref{mspsensplot} shows a comparison between the sensitivities of the
mid-latitude and low-latitude HTRU survey and the PMPS for a
DM of 100~cm$^{-3}$pc. Although the mid-latitude HTRU survey has only
one quarter the observing time of the PMPS, it is significantly
more sensitive to short period pulsars at large DMs. This is due
to a combination of the faster sampling rate, narrower frequency channels,
slightly increased total bandwidth and multi-bit sampling of our survey.

Of the new discoveries, 9 pulsars lie within the sky coverage of the PMPS.
We have re-analysed archival data for each of these pulsars and the results are presented in Table \ref{pmreanaly}.
Four of these pulsars are MSPs with high DMs and are included on Figure~\ref{mspsensplot}.
It can be seen that all bar one fall below the (nominal) PMPS detection threshold.
In fact, re-examination of the PMPS archival data shows that both
PSR~J1708--35 and J1801--32 are visible but that the pulse broadening
from the 3~MHz channel bandwidth makes them difficult to distinguish
from sinusoidal interference in a blind search.

Of the remaining five pulsars, three are seen in the archival data with low SNR and none were selected for re-observation in the PMPS.
Although the pulsar positions are not well known at this time, the nominal position of each falls more than 5 arc-minutes from the beam centre of the PMPS observation.
Therefore we suggest that these pulsars fall in the regions between beam centres where the PMPS sensitivity is lowest.
Two pulsars appear to be below the detection limit of the PMPS including the newly discovered
magnetar PSR J1622--4950, which shows extreme time variability \cite{hitrun_magnetar}.

\begin{table}
\caption{
Pulsars discovered in the HTRU survey for which archival data exists in the PMPS at the position of the pulsar.
The unique PMPS grid identification number (see \citealp{mlc+01} for details) for the processed beam, the MJD of the observation and the SNR and periodicity if the pulsar was detected.
\label{pmreanaly}
}
\begin{center}
{\footnotesize \tt 
\begin{tabular}{lllrl}
\hline
PSR J      &  Grid ID  & MJD      & SNR  & Period (s)    \\
\hline
1125$-$58   & G4707513 & 51296.42 & --   & --        \\
1622$-$4950 & G4887499 & 50686.25 & --   & --        \\
1625$-$49   & G4891500 & 50686.23 & 7.9  & 0.3558530 \\
1701$-$44   & G4924493 & 51152.94 & 6.4  & 0.7555360 \\
1705$-$43   & G4928493 & 51457.10 & 10.4 & 0.2225614 \\
1708$-$35   & G4959516 & 51718.68 & 17.4 & 0.0045052 \\
1754$-$24   & G5021504 & 50843.07 & --   & --        \\
1801$-$32   & G4995477 & 52080.69 & 14.4 & 0.0075344 \\
1811$-$24   & G5030488 & 51245.84 & --   & --        \\
\hline
\end{tabular}
}
\end{center}
\end{table}

\section{Conclusion}
We have described the system configuration and initial discoveries of a new pulsar survey using the Parkes radio telescope.
At the time of writing, a total of 250 pulsars have been detected, of which 27 are new discoveries.
The high time and frequency resolution of our digital backend system leads to increased sensitivity to short-period, high-DM pulsars compared to previous surveys.
Indeed, 5 of our new discoveries are MSPs at high DM, all of which lie within previously surveyed regions of sky.

\section{Acknowledgements}
The Parkes Observatory is part of the Australia Telescope which is funded by the Commonwealth of Australia for operation as a National Facility managed by CSIRO.
The HYDRA supercomputer at the JBCA was supported by a grant from the UK Science and Technology Facilities Council.
S.B. gratefully acknowledges the support of STFC in his PhD studentship.

\bibliographystyle{mnras}
\bibliography{journals,myrefs,modrefs,psrrefs,crossrefs}

\begin{thebibliography}{}

\bibitem[\protect\citeauthoryear{{Archibald} et~al.}{{Archibald}
  et~al.}{2009}]{asr+09}
{Archibald} A.~M. et~al., 2009, Science, 324, 1411

\bibitem[\protect\citeauthoryear{{Burgay} et~al.}{{Burgay}
  et~al.}{2003}]{bdp+03}
{Burgay} M. et~al., 2003, Nature, 426, 531

\bibitem[\protect\citeauthoryear{{Burke-Spolaor} \& {Bailes}}{{Burke-Spolaor}
  \& {Bailes}}{2010}]{bb10}
{Burke-Spolaor} S.,  {Bailes} M., 2010, MNRAS, 402, 855

\bibitem[\protect\citeauthoryear{Cordes \& Chernoff}{Cordes \&
  Chernoff}{1997}]{cc97}
Cordes J.~M.,  Chernoff D.~F., 1997, ApJ, 482, 971

\bibitem[\protect\citeauthoryear{{Cordes} \& {Lazio}}{{Cordes} \&
  {Lazio}}{2002}]{cl02}
{Cordes} J.~M.,  {Lazio} T.~J.~W., preprint (arXiv:astro-ph/0207156)

\bibitem[\protect\citeauthoryear{{Eatough}, {Keane}, \& {Lyne}}{{Eatough}
  et~al.}{2009}]{ekl09}
{Eatough} R.~P., {Keane} E.~F.,  {Lyne} A.~G., 2009, MNRAS, 395, 410

\bibitem[\protect\citeauthoryear{{Eatough} et~al.}{{Eatough}
  et~al.}{2010}]{emk+10}
{Eatough} R.~P., {Molkenthin} N., {Kramer} M., {Noutsos} A., {Keith} M.~J.,
  {Stappers} B.~W.,  {Lyne} A.~G., 2010, MNRAS, submitted

\bibitem[\protect\citeauthoryear{{Edwards} et~al.}{{Edwards}
  et~al.}{2001}]{ebvb01}
{Edwards} R.~T., {Bailes} M., {van Straten} W.,  {Britton} M.~C., 2001, MNRAS,
  326, 358

\bibitem[\protect\citeauthoryear{{Haslam} et~al.}{{Haslam}
  et~al.}{1981}]{hks+81}
{Haslam} C.~G.~T., {Klein} U., {Salter} C.~J., {Stoffel} H., {Wilson} W.~E.,
  {Cleary} M.~N., {Cooke} D.~J.,  {Thomasson} P., 1981, A\&A, 100, 209

\bibitem[\protect\citeauthoryear{{Hessels} et~al.}{{Hessels}
  et~al.}{2006}]{hrs+06}
{Hessels} J.~W.~T., {Ransom} S.~M., {Stairs} I.~H., {Freire} P.~C.~C., {Kaspi}
  V.~M.,  {Camilo} F., 2006, Science, 311, 1901

\bibitem[\protect\citeauthoryear{Hewish et~al.}{Hewish et~al.}{1968}]{hbp+68}
Hewish A., Bell S.~J., Pilkington J.~D.~H., Scott P.~F.,  Collins R.~A., 1968,
  Nature, 217, 709

\bibitem[\protect\citeauthoryear{{Hobbs} et~al.}{{Hobbs} et~al.}{2009}]{hbb+09}
{Hobbs} G.~B. et~al., 2009, Publ. Astr. Soc. Aust., 26, 103

\bibitem[\protect\citeauthoryear{{Hotan}, {van Straten}, \&
  {Manchester}}{{Hotan} et~al.}{2004}]{hvm04}
{Hotan} A.~W., {van Straten} W.,  {Manchester} R.~N., 2004, Proc. Astr. Soc.
  Aust., 21, 302

\bibitem[\protect\citeauthoryear{{Jacoby} et~al.}{{Jacoby}
  et~al.}{2009}]{jbo+09}
{Jacoby} B.~A., {Bailes} M., {Ord} S.~M., {Edwards} R.~T.,  {Kulkarni} S.~R.,
  2009, ApJ, 699, 2009

\bibitem[\protect\citeauthoryear{{Keith} et~al.}{{Keith} et~al.}{2009}]{kel+09}
{Keith} M.~J., {Eatough} R.~P., {Lyne} A.~G., {Kramer} M., {Possenti} A.,
  {Camilo} F.,  {Manchester} R.~N., 2009, MNRAS, 395, 837

\bibitem[\protect\citeauthoryear{{Keith} et~al.}{{Keith} et~al.}{2008}]{kjk+08}
{Keith} M.~J., {Johnston} S., {Kramer} M., {Weltevrede} P., {Watters} K.~P.,
  {Stappers} B.~W., 2008, MNRAS, 389, 1881

\bibitem[\protect\citeauthoryear{{Kramer} et~al.}{{Kramer}
  et~al.}{2006}]{ksm+06}
{Kramer} M. et~al., 2006, Science, 314, 97

\bibitem[\protect\citeauthoryear{{Levin} et~al.}{{Levin}
  et~al.}{2010}]{hitrun_magnetar}
{Levin} L. et~al., 2010, Science, submitted

\bibitem[\protect\citeauthoryear{{Lorimer} et~al.}{{Lorimer}
  et~al.}{2007}]{lbm+07}
{Lorimer} D.~R., {Bailes} M., {McLaughlin} M.~A., {Narkevic} D.~J.,  {Crawford}
  F., 2007, Science, 318, 777

\bibitem[\protect\citeauthoryear{{Lorimer} et~al.}{{Lorimer}
  et~al.}{2006}]{lfl+06}
{Lorimer} D.~R. et~al., 2006, MNRAS, 372, 777

\bibitem[\protect\citeauthoryear{Lorimer \& Kramer}{Lorimer \&
  Kramer}{2005}]{lk05}
Lorimer D.~R.,  Kramer M., 2005, Handbook of Pulsar Astronomy.
\newblock Cambridge University Press

\bibitem[\protect\citeauthoryear{Lyne et~al.}{Lyne et~al.}{2004}]{lbk+04}
Lyne A.~G. et~al., 2004, Science, 303, 1153

\bibitem[\protect\citeauthoryear{Lyne et~al.}{Lyne et~al.}{1998}]{lml+98}
Lyne A.~G. et~al., 1998, MNRAS, 295, 743

\bibitem[\protect\citeauthoryear{Manchester et~al.}{Manchester
  et~al.}{2001}]{mlc+01}
Manchester R.~N. et~al., 2001, MNRAS, 328, 17

\bibitem[\protect\citeauthoryear{Manchester et~al.}{Manchester
  et~al.}{1996}]{mld+96}
Manchester R.~N. et~al., 1996, MNRAS, 279, 1235

\bibitem[\protect\citeauthoryear{{McLaughlin} et~al.}{{McLaughlin}
  et~al.}{2006}]{mll+06}
{McLaughlin} M.~A. et~al., 2006, Nature, 439, 817

\bibitem[\protect\citeauthoryear{McMahon}{McMahon}{2008}]{mcm08}
McMahon P., 2008, Master's thesis, University of Cape Town

\bibitem[\protect\citeauthoryear{Staveley-Smith et~al.}{Staveley-Smith
  et~al.}{1996}]{swb+96}
Staveley-Smith L. et~al., 1996, Proc. Astr. Soc. Aust., 13, 243

\bibitem[\protect\citeauthoryear{Taylor}{Taylor}{1974}]{tay74}
Taylor J.~H., 1974, A\&AS, 15, 367

\bibitem[\protect\citeauthoryear{{Verbiest} et~al.}{{Verbiest}
  et~al.}{2009}]{vbc+09}
{Verbiest} J.~P.~W. et~al., 2009, MNRAS, 400, 951

\bibitem[\protect\citeauthoryear{{Weltevrede} et~al.}{{Weltevrede}
  et~al.}{2010}]{waa+10}
{Weltevrede} P. et~al., 2010, ApJ, 708, 1426

\bibitem[\protect\citeauthoryear{{Yusifov} \& {K{\"u}{\c c}{\"u}k}}{{Yusifov}
  \& {K{\"u}{\c c}{\"u}k}}{2004}]{yk04}
{Yusifov} I.,  {K{\"u}{\c c}{\"u}k} I., 2004, A\&A, 422, 545

\end{thebibliography}

\end{document}